\documentstyle[12pt]{article}
\begin{document}
\begin{center}
\begin{Large}
\begin{bf}
Why We Already Know that Antihydrogen\\
is Almost Certainly NOT Going to Fall ``Up'' 
\end{bf}
\end{Large}
\vskip 1cm

\begin{large}
Scott Menary\footnote{menary@yorku.ca}\\ 
Department of Physics \& Astronomy\\
York University\\ 
Toronto, Canada
\end{large}
\end{center}
\bigskip

\begin{quote}
``In order for the light to shine so brightly, the darkness must be present.''\\
- Sir Francis Bacon (1561 -- 1626)
\smallskip

``We lose a great deal when we lose the sense and feeling for the sun. When all has been said, 
the adventure of the sun is the great natural drama by which we live, and not to have joy in 
it and awe of it, not to share in it, is to close a dull door on nature's sustaining and 
poetic spirit.''\\
- Henry Beston (1888 -- 1968)
\end{quote}
\bigskip

The ALPHA collaboration (of which I am a member) has made great strides recently in
trapping antihydrogen\cite{trap,long} and starting down the path of making spectroscopic 
measurements\cite{spec}. The primary goal of the experiment is to test CPT invariance but
there is also interest in testing another fundamental issue -- the gravitational interaction
between matter and antimatter (the so-called question of ``antigravity''). As well as the other
antihydrogen trapping experiments -- ASACUSA\cite{asacusa} and ATRAP\cite{atrap} -- 
there is also a new experiment in the
Antiproton Decelerator hall at CERN called AEGIS\cite{aegis} which is dedicated
to testing the gravitional interaction between antihydrogen and the Earth. It was claimed 
in \cite{lykken} that there ``is no compelling evidence or theoretical reason to rule out such
a difference (i.e., between $g$ and $\bar{g}$) at the 1\% level.'' I argue in this short paper
that bending of light by the sun provides a more stringent limit than this\footnote{After
having written this note I was referred to a paper by Nieto and Goldman\cite{nieto} where
there is a brief mention of a test of this type.}. 

It was pointed out in \cite{wag} something that is so obvious and simple that I had one of
those ``why didn't I think of that?'' moments. Let's assume that matter and antimatter repel
gravitationally. What happens to a particle which is its own antiparticle when it is in the 
gravitational field of, say, the Earth? The example given in \cite{wag} of a particle 
which is its own antiparticle is the $\pi^0$ (which, you'll recall, has a quark content of 
$1/\sqrt{2}[\vert u>\vert \bar{u}>+\vert d>\vert \bar{d}>]$). Other examples include Positronium,
the $J/\psi$ ($c\bar{c}$), and the $\Upsilon$ ($b\bar{b}$). Presumably any of these objects
would just float where they were since the attractive and repulsive forces would be
equal and opposite (I'll get back to the question of the forces being equal, which really is an 
experimental question). That is, they wouldn't ``fall'' up or down but just stay put.

This is purely on the level of a gedankenexperiment at this point since the $\pi^0$ lifetime is
$8.5\times 10^{-17}$ s and the long-lived state of positronium has a lifetime of 142 ns. The
heavy quark-antiquark states have lifetimes on the order of $10^{-20}$ s. So 
nobody is going to do this experiment any time soon (or ever) with these objects. But there is
one particle which is its own antiparticle that has essentially an infinite lifetime -- the
photon. Now, given the reasoning of the previous paragraph, a photon's trajectory should be 
completely unaffected by the presence of matter in the region through which it is traveling.
And, of course, we know that this isn't true. The deflection of light by the sun was the
measurement that made Einstein and his General Theory of Relativity (GR) famous. There is also
gravitational lensing (which, of course, is essentially the same effect). This shows
{\it experimentally} that antimatter is attracted to matter. Recall from QED that the photon
isn't really a point particle but is more like a cloud of $e^+e^-$ pairs (plus the occasional
$\mu^+\mu^-$ pair). This really illustrates the point that it is equal amounts particle and 
antiparticle. So there is no other conclusion that can be reached -- matter and antimatter
attract gravitionally. 

The argument given above assumes that the strength of the gravitational 
interaction between matter and
antimatter is the same as for that between matter and matter. I should say that,
technically speaking, if this isn't the case then this again opens up the possibility that 
matter and antimatter repel. The only requirement then is 
that the strength of the repulsion is weaker than the strength of the attractive
matter-matter interaction. That is, the sun is attracting the ``matter part'' of the photon
with a greater magnitude than it is repeling the ``antimatter part'' so the photon would still
bend around the sun, just not as much as it would if the two couplings were the same. 
We'll see that the repulsive matter-antimatter coupling is, at most, 0.5\%
that of the matter-matter coupling, which seems to strain our ability to observe it. And
while I am the first one to agree that physics is an experimental science, it is hard to 
imagine what theoretical contortions are necessary in order to explain such an effect.

The main point is, irrespective of whether the matter-antimatter gravitational interaction is 
repulsive 
or attractive, if the strength of the matter-matter gravitational interaction (which goes like $G$) 
is different from the matter-antimatter interaction (say it goes like $\bar{G}$ and recall, 
from the arguments given above, that it can only be weaker if it is repulsive),
then the deflection of light by the sun will differ from the predictions of GR. 
That was a long sentence. If $G$ = $\bar{G}$
and the matter-antimatter interaction is attractive then the deflection should be exactly as
predicted by GR. So the idea is to compare measurements of the deflection of light by the
sun with the predictions of GR. Just such a comparison is described in \cite{will}. 
Modern measurements
using radio waves are actually pretty precise. Any potential difference from GR (i.e., extensions
to GR or different types of metric theories) is encorporated in the parameter $\gamma$. For GR,
$\gamma$ = 1. The result as of 2004 from 87 VLBI sites using almost 2 million VLBI observations 
of 541 radio sources is $\gamma - 1$ = $(-1.7\pm 4.5)\times 10^{-4}$. So, in other words, if
$\bar{G}\ne G$ and the matter-antimatter gravitional interaction is repulsive, 
then $\bar{G}$ is, {\it at most}, about 0.5\% of $G$. I'm not sure if the constraint
might actually be twice as strong if the matter-antimatter interaction is replusive but I think
that is beside the point, this is probably at least the upper limit. 

To summarize, a repulsive matter-antimatter gravitational force 
is constrained by {\it observation} to be, at most, 0.5\% as strong as the attractive
matter-matter gravitational force.
Similarly, even if the matter-antimatter gravitational interaction is
attractive, it cannot differ in strength (either bigger or smaller) 
from the matter-matter gravitational interaction by more than about 0.5\%.
It seems to me, unfortunately, that this makes the measurement of such an effect in the
antihydrogen experiments
just that much more difficult since we now need a sensitivity which is at least a factor of 2
better than what we'd been assuming. 
\bigskip

\begin{large}
\noindent {\bf Afterword}
\end{large}
\bigskip

I sent this idea to Joe Lykken, a theoretical physicist working at Fermilab who is one of the authors 
of \cite{lykken}, and he replied\cite{priv} 
with the following two possible ways in which one can theoretically get around the 0.5\% 
limit\footnote{These arguments can also be found in \cite{alves}}.
\begin{itemize}
\item[1.] The simplest way to get a matter-antimatter ``gravity" asymmetry is not to alter 
gravity itself, but
rather to introduce new ``$5^{\rm th}$ forces'' carried by very light scalars and/or gauge bosons, and
couple to some combination of baryon or lepton number. In that case [the above] argument applies,
but the effect on photons is suppressed by at least a QED loop factor ~100, since the photon does
not carry B or L.
\item[2.] In ordinary GR the photon only couples to the trace of the energy-momentum tensor. If some
modification of GR creates a matter-antimatter gravity asymmetry, then I can imagine that it
might simply vanish when you trace the modified version of the Einstein equations, so it
doesn't affect photons at leading order.
\end{itemize}

Even given this the observations articulated in this note may still be useful. 
Say ``antigravity'' is observed at the 1\% level
in measurements performed using antihydrogen. Then this analysis could be used to put constraints
on the 
types of models proposed involving antigravity since any such model would have to accomodate both the 
bending of light by the sun, in agreement with GR to 0.5\%, and 
a difference between $g$ and $\bar{g}$ at the 1\% level found in the antihydrogen experiment. 
\bigskip

\begin{large}
\noindent {\bf Acknowledgements}
\end{large}
\bigskip

I would like to thank Makoto Fujiwara, Joe Lykken, Dan Kaplan, Chanpreet Amole, and Andrea Capra
for their comments and Clifford Will for pointing me to the newest results on the bending of light
by the sun.

\end{document}